\documentclass[prl,twocolumn,showpacs]{revtex4}

\usepackage[dvips]{graphicx}
\usepackage{dcolumn}% Align table columns on decimal point
\usepackage{bm}% bold math
\usepackage{amsmath}
\usepackage{ulem}

\begin{document}

% \textit{Physical Review Letters}
% \preprint{APS/123-QED}

\title{Pressure Dependence of the Magnetic Anisotropy in the "Single-Molecule Magnet" [Mn$_{4}$O$_{3}$Br(OAc)$_{3}$(dbm)$_{3}$]}

\author{Andreas Sieber}
\author{Gr\'{e}gory Chaboussant}
\author{Roland Bircher}%
\author{Colette Boskovic}%
\author{Hans U. G\"{u}del}%
\affiliation{%\\
Department of Chemistry and Biochemistry, University of Bern,
Freiestrasse 3, CH-3000 Bern 9, Switzerland}%
\author{George Christou}
\affiliation{\\
Department of Chemistry, University of Florida, Gainesville, FL, 32611, USA}%
\author{Hannu Mutka}
\affiliation{%\\
Institut Laue-Langevin, 6 rue Jules Horowitz, BP 156, 38042 Grenoble Cedex 9, France}%

\date{\today}

\begin{abstract}
The anisotropy splitting in the ground state of the single-molecule magnet
$\rm[Mn_{4}O_{3}Br(OAc)_{3}(dbm)_{3}]$ is studied by inelastic neutron scattering as
a function of hydrostatic pressure. This allows a tuning of the anisotropy and thus
the energy barrier for slow magnetisation relaxation at low temperatures. The value
of the negative axial anisotropy parameter $D_{\rm cluster}$ changes from -0.0627(1)
meV at ambient to -0.0603(3) meV at 12 kbar pressure, and in the same pressure range
the height of the energy barrier between up and down spins is reduced from 1.260(5)
meV to 1.213(9) meV. Since the $\rm Mn-Br$ bond is significantly softer and thus
more compressible than the $\rm Mn-O$ bonds, pressure induces a tilt of the single
ion Mn$^{3+}$ anisotropy axes, resulting in the net reduction of the axial cluster
anisotropy.
\end{abstract}


\maketitle

Single-molecule magnets (SMM) are presently the focus of a very intense research
activity. SMM are molecules containing a finite number of exchange coupled magnetic
ions, and they exhibit phenomena such as slow relaxation and quantum tunneling of
the magnetisation at low temperatures \cite{Gatteschi_rev}. They are the smallest
known units that are potentially capable of storing a bit of information at
cryogenic temperatures. An easy axis type magnetic anisotropy is an essential
prerequisite for an energy barrier between up and down spins and thus for slow
relaxation. The height of this barrier is determined by both the ground state
\textit{S} value and the size of the negative \textit{D} value in the axial spin
Hamiltonian

\begin{equation}
\label{eq:barrier} \hat {H}_{axial}\, = \,D\left[ {\hat {S}_z2 - \frac{1}{3}S\left(
{S + 1} \right)} \right]
\end{equation}

\begin{figure}
\includegraphics[width=80mm]{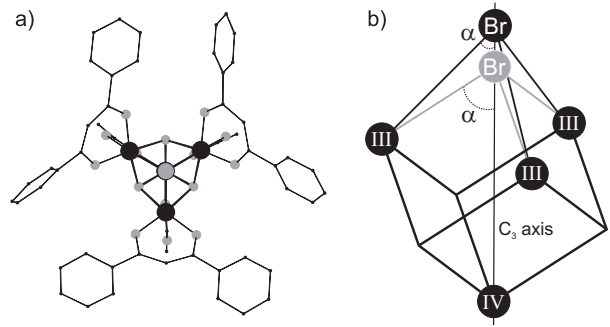}
\caption{a) Molecular structure of the title complex Mn$_{4}$, from Ref.\
\cite{Structure}. View along the approximate C$_{3}$ axis. For clarity the H atoms
are omitted. Mn$^{3+}$ ions are drawn as large black spheres, C and O atoms as small
black and grey spheres, respectively. The large grey sphere represents the Br$^{-}$
ion, which obscures the Mn$^{4+}$ ion just behind. b) Schematic view of the core of
Mn$_{4}$, with III and IV representing Mn$^{3+}$ and Mn$^{4+}$, respectively. The
black and grey positions of the Br$^{-}$ ion schematically represent the situation
at ambient and high external pressure, respectively. \label{fig:cluster}}
\end{figure}

For even and odd \textit{S} values the barrier height is given by $|D|S^{2}$ and
$|D|(S^{2}-\frac{1}{4})$, respectively. Chemists have been able to assemble numerous
spin clusters which show SMM features at the very lowest temperatures. But the
number of examples with blocking temperatures above 1 K is still rather limited.
Among them is a family of tetranuclear manganese clusters with general formula
$\rm[Mn_{4}O_{3}X(OAc)_{3}(dbm)_{3}]$, where OAc$^{-}$ is the acetate ion and
dbm$^{-}$ is the anion of dibenzoylmethane. They all exhibit SMM behavior with an
energy barrier of the order of 1.25 meV \cite{Hanspeter}. Here we report the first
direct spectroscopic determination of the anisotropy splitting in a SMM under
hydrostatic pressure. The molecule $\rm[Mn_{4}O_{3}Br(OAc)_{3}(dbm)_{3}]$ (Mn$_{4}$)
belongs to the above family, and its molecular structure is shown in Figure
\ref{fig:cluster}a.

The molecule has a $\rm [Mn^{4+}(Mn^{3+})_{3}(\mu_{3}-O)_{3}(\mu_{3}-Br)]^{6+}$ core
with a distorted cubane geometry, which is schematically depicted in Figure
\ref{fig:cluster}. The molecular point symmetry is approximately C$_{3v}$ with the
C$_{3}$ axis passing through the Mn$^{4+}$ and Br$^{-}$ ions (Figure
\ref{fig:cluster}b) \cite{Structure}. We correlate the pressure dependence of the
anisotropy splitting with pressure induced changes in the structure and identify the
dominant terms and factors which govern the anisotropy splitting and thus the
barrier height.

Inelastic neutron scattering (INS) is the most direct technique to measure
anisotropy splittings in SMMs in the absence of an external magnetic field. Among
others, anisotropy parameters have thus been obtained for the prototype SMMs
Mn$_{12}$-acetate \cite{Mirebeau} and $\rm [Fe_{8}O_{2}(OH)_{12}(tacn)_{6}]^{8+}$
\cite{Fe8} as well as four members of the Mn$_{4}$ cubane family including the title
compound \cite{Hanspeter}.

\begin{figure}[h!]
\includegraphics[width=80mm]{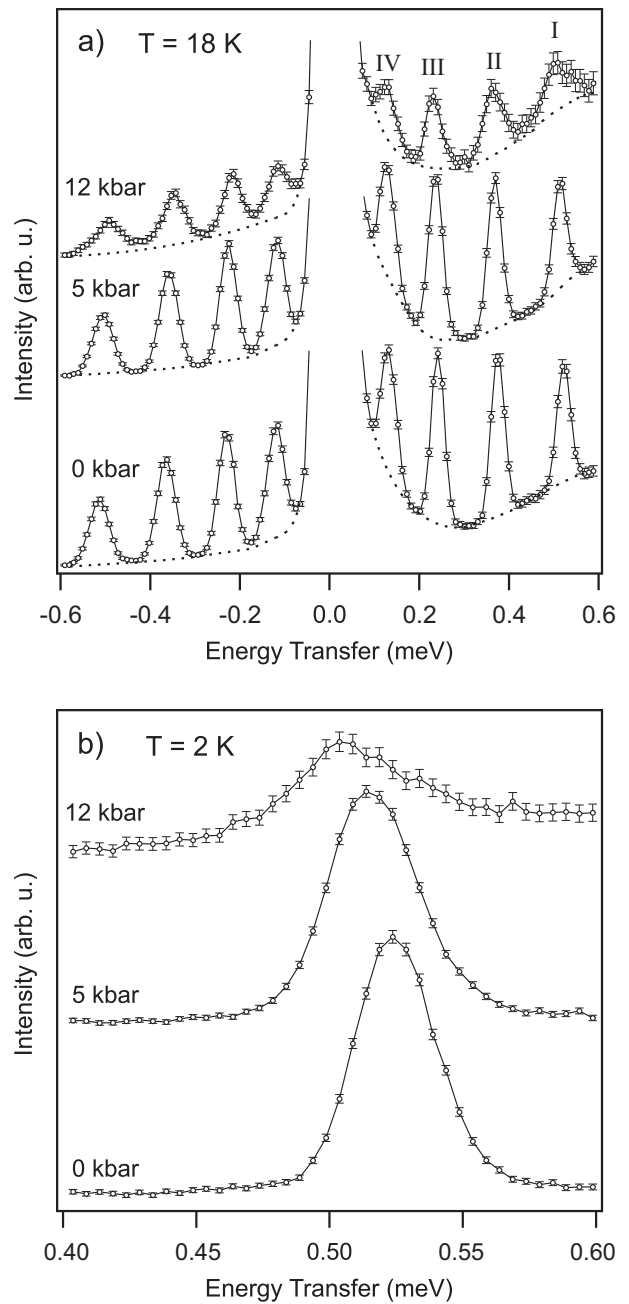}
\caption{a) INS spectra at 18 K of  a polycrystalline sample of partially deuterated
[Mn$_{4}$O$_{3}$Br(d$_{3}$-OAc)$_{3}$(dbm)$_{3}$] recorded on IN5 with an incident
wavelength $\lambda_{i} = 7.5$ $\rm\AA$ for pressures \textit{p} = 0, 5 and 12 kbar.
The spectra correspond to the sum of all the scattering angles. The labeling of the
peaks corresponds to Table \ref{table:peaks} and Figure \ref{fig:barrier}. The full
lines connect adjacent data points. The dotted lines illustrate the background which
was approximated by polynomials and subtracted from the data. b) Pressure dependence
of peak I at 2 K.} \label{fig:data}
\end{figure}

The present measurements were carried out on a partially deuterated sample with
composition $\rm[Mn_{4}O_{3}Br(d_{3}$-$\rm OAc)_{3}(dbm)_{3}]$ using the
time-of-flight spectrometer IN5 at the Institut Laue Langevin (ILL) in Grenoble. The
sample was prepared according to Ref.\ \cite{Synthesis}. For pressures of 0 kbar, 3
kbar and 5 kbar about 2 grams of polycrystalline sample placed in a standard ILL
continuously loaded high-pressure cell with He as the pressure transmitting medium
were used. For 12 kbar the standard ILL high-pressure clamped cell was employed with
about 0.3 grams of sample. Neutron wavelengths of 7.5 $\rm\AA$ (0 kbar to 12 kbar)
and 8.5 $\rm\AA$ (0 kbar to 5 kbar) were used, corresponding to instrumental
resolutions of 32 $\mu$eV and 19 $\mu$eV, respectively. The data treatment involved
the calibration of the detectors by means of a spectrum of vanadium metal.

Experimental results for 0 kbar, 5 kbar and 12 kbar at 18 K are shown in Figure
\ref{fig:data}a. At this temperature all the ground state levels have some
population. At all pressures four well resolved inelastic peaks, labelled I to IV,
are observed on both the energy loss and gain side, corresponding to positive and
negative energy transfers in Figure \ref{fig:data}a, respectively. The 12 kbar peaks
are slightly inhomogeneously broadened. At 2 K only peak I is observed, and this is
shown on an expanded energy scale in Figure \ref{fig:data}b. A decrease of the peak
energy with pressure is evident. An analysis using Gaussian fits to the background
corrected data yields the peak positions in Table \ref{table:peaks}. The data at
ambient pressure are in good agreement with those reported in Ref.\
\cite{Hanspeter}.

Antiferromagnetic exchange interactions between the three Mn$^{3+}$ (\textit{S} = 2)
ions and the Mn$^{4+}$ (\textit{S} = $\frac{3}{2}$) ion dominate the coupling in
Mn$_{4}$, thus leading to a \textit{S} = $\frac{9}{2}$ cluster ground state, with
the first excited \textit{S} = $\frac{7}{2}$ state at about 22 meV and thus outside
the range of our experiment. The trigonal symmetry of the Mn$_{4}$ molecules in the
crystal structure of the title compound is slightly distorted, as can be seen in
Figure \ref{fig:cluster}a, with an actual point group symmetry C$_{1}$. Including a
higher order term the appropriate spin Hamiltonian to account for the splitting of
the \textit{S} = $\frac{9}{2}$ ground state is thus given by

\begin{equation}
\label{eq:Hamiltonian} \hat {H}_{\rm aniso}\, = \,D\left[ {\hat {S}_z2 -
\frac{1}{3}S\left( {S + 1} \right)} \right] + B_40 \hat {O}_40 + E\left( {\hat
{S}_x2 - \hat {S}_y2} \right)
\end{equation}
where $\hat {O}_40 \, = \,35\hat {S}_z4 - \left[ {30S(S + 1) - 25} \right]\hat
{S}_z2 - 6S(S + 1) + 3\hat {S}^2\left( {S + 1} \right)2$.

\begin{figure}
\includegraphics[width=80mm]{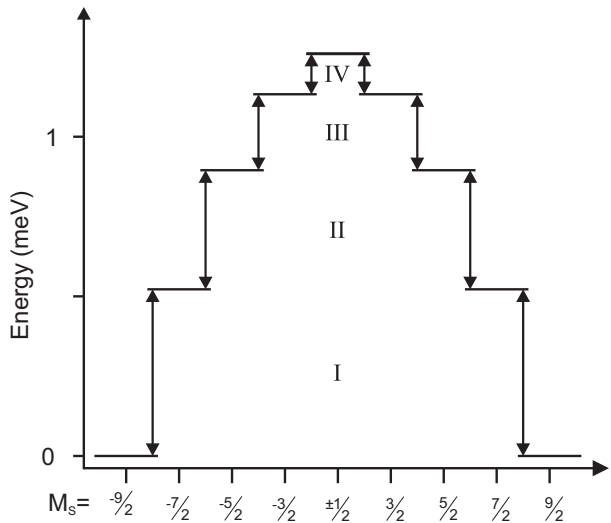}
\caption{Calculated ansisotropy splitting in the $S = \frac{9}{2}$ ground state of
the title compound with the parameter values \textit{D} = -0.0627 meV, $B_{4}^{0} =
-6.2\cdot 10^{-6}$ meV and $|E| = 1.9\cdot 10^{-3}$ meV. The double arrows
correspond to the allowed $\Delta M_{S} = \pm 1$ INS transitions.
\label{fig:barrier}}
\end{figure}

From the data in Ref.\ \cite{Hanspeter} the following parameter values at 18 K were
determined: \textit{D} = -0.062 meV, $B_{4}^{0}$ = $-6.3\cdot10^{-6}$ meV and $|E|$
= $2.1\cdot10^{-3}$ meV. The first term in Eq.\ \ref{eq:Hamiltonian} is the leading
term, and thus $M_{S}$ remains a reasonably good quantum number. The splitting
pattern with the above parameters is shown in Figure \ref{fig:barrier}. Magnetic
neutron scattering theory leads to the selection rules $\Delta M_{S} = 0, \pm 1$ for
INS, i.\ e.\ transitions between adjacent levels are allowed, see the arrows in
Figure \ref{fig:barrier}. We can thus immediately assign the observed INS bands I to
IV in Figure \ref{fig:data} as given in the second column of Table
\ref{table:peaks}. Fitting the eigenvalues of Eq.\ \ref{eq:Hamiltonian} to the
observed band energies yields the parameter values at the bottom of Table
\ref{table:peaks}. Both $|B_{4}^{0}|$ and $|E|$ are much smaller than $|D|$, but
they are essential for a proper description, and they are responsible for the
deviations from a regular spacing of the peaks in Figure \ref{fig:data}a. The
parameter values at ambient pressure are the same within experimental accuracy as
those derived from the data in Ref.\ \cite{Hanspeter}. The negative \textit{D} is
significantly pressure dependent, its value decreasing linearly by 3.8\% between
ambient pressure and 12 kbar. This leads to a reduction of the energy barrier, i.\
e.\ the energy difference between the $M_{S} = \pm\frac{1}{2}$ and $M_{S} =
\pm\frac{9}{2}$ levels, from 1.260(5) meV at ambient to 1.213(9) meV at 12 kbar
pressure, respectively. The pressure dependence of $B_{4}^{0}$ and \textit{E} is too
small to be determined by our experiment.

\begin{table*}
\begin{tabular}{cccccccccccccc}\\
\hline \vspace{-8pt}\\
& & \multicolumn{12}{c}{energy (meV)}\\
\vspace{1pt} & & \hspace{2pt} & \multicolumn{2}{c}{\textit{p} = 0 kbar} &
\hspace{2pt} & \multicolumn{2}{c}{\textit{p} = 3 kbar}
& \hspace{2pt} & \multicolumn{2}{c}{\textit{p} = 5 kbar} & \hspace{2pt} & \multicolumn{2}{c}{\textit{p} = 12 kbar}\\
\cline{4-5} \cline{7-8} \cline{10-11} \cline{13-14} \vspace{-9pt}\\
label & transition & & exp. & calc. & & exp. & calc. & & exp. & calc. & & exp. & calc. \\
\hline \vspace{-8pt}\\
I & $\pm\frac{9}{2}\leftrightarrow\pm\frac{7}{2}$ & & 0.522(1) & 0.522 & & 0.517(1) & 0.517 & & 0.514(1) & 0.514 & & 0.503(1) & 0.503 \vspace{2pt}\\
II & $\pm\frac{7}{2}\leftrightarrow\pm\frac{5}{2}$ & & 0.373(1) & 0.373 & & 0.369(1) & 0.369 & & 0.366(1) & 0.367 & & 0.359(3) & 0.359 \vspace{2pt}\\
III & $\pm\frac{5}{2}\leftrightarrow\pm\frac{3}{2}$ & & 0.238(1) & 0.238 & & 0.235(1) & 0.235 & & 0.233(1) & 0.233 & & 0.227(2) & 0.228 \vspace{2pt}\\
IV & $\pm\frac{3}{2}\leftrightarrow\pm\frac{1}{2}$ & & 0.127(2) & 0.127 & & 0.126(1) & 0.126 & & 0.125(1) & 0.125 & & 0.124(3) & 0.123 \vspace{2pt}\\
\hline \vspace{-8pt}\\
\multicolumn{2}{c}{\textit{D}} & & \multicolumn{2}{c}{-0.0627(1)} & &
\multicolumn{2}{c}{-0.0620(1)} & &
\multicolumn{2}{c}{-0.0616(1)} & & \multicolumn{2}{c}{-0.0603(3)} \\
\multicolumn{2}{c}{$|E|$} & & \multicolumn{2}{c}{$1.90(1)\cdot10^{-3}$} & &
\multicolumn{2}{c}{$1.90\cdot10^{-3}$} & & \multicolumn{2}{c}{$1.90\cdot10^{-3}$} &
&
\multicolumn{2}{c}{$1.90\cdot10^{-3}$} \\
\multicolumn{2}{c}{$B_{4}^{0}$} & & \multicolumn{2}{c}{$-6.2(2)\cdot10^{-6}$} & &
\multicolumn{2}{c}{$-6.2\cdot10^{-6}$} & & \multicolumn{2}{c}{$-6.2\cdot10^{-6}$} &
&
\multicolumn{2}{c}{$-6.2\cdot10^{-6}$}\vspace{2pt}\\
\hline
\end{tabular}
\caption{\label{table:peaks} Experimental INS peak positions and calculated
transition energies using Eq.\ \ref{eq:Hamiltonian} as a function of hydrostatic
pressure. The labeling corresponds to Figure \ref{fig:data}, and the assignments in
the second column refer to Figure \ref{fig:barrier}. Parameter values for each
pressure are given at the bottom. The pressure dependence of $|E|$ and $B_{4}^{0}$
is too small to be determined in this experiment, and the values determined at
ambient pressure were used throughout.}
\end{table*}

The only pressure experiments on single-molecule magnets reported in the literature
are for Mn$_{12}$-acetate. From the pressure dependence of the low temperature
magnetisation it was concluded that pressure produces a geometrical molecular isomer
of Mn$_{12}$-acetate with significantly faster relaxation of the magnetisation
\cite{Mn12pressurePRB}. On the other hand, changes in the position of the steps in
the hysteresis of Mn$_{12}$-acetate under pressure were ascribed to an increase of
the axial anisotropy splitting with pressure \cite{Mn12pressure}. Our experimental
finding for Mn$_{4}$ that the axial anisotropy splitting is decreasing with pressure
is unambiguous. With the following simplified model we account for this decrease by
correlating it with the expected structural changes of the molecule under pressure.
We assume an isotropic compressibility for the core defined by the three Mn$^{3+}$
ions and the Mn$^{4+}$ ion in Figure \ref{fig:cluster}b. All the metal-ligand bonds
in this core are either Mn$^{3+}-$O or Mn$^{4+}-$O bonds, and taking average linear
compressibilities (\textit{d}) from the literature we calculate \textit{d}(12
kbar)/\textit{d}(ambient) = 0.9975 \cite{compressibility}. The Mn$^{3+}-$Br bonds in
Figure \ref{fig:cluster} are significantly softer and more compressible than the
Mn$-$O bonds. A ratio of force constants
\textit{k}(Mn$^{3+}-$O)/\textit{k}(Mn$^{3+}-$Br) = 4.2 is obtained from literature
values based on Raman experiments \cite{force1,force2}. In terms of compressibility
we thus calculate a ratio \textit{d}(12 kbar)/\textit{d}(ambient) = 0.993 for the
Mn$^{3+}-$Br$^{-}$ bonds. The net effect of pressure in this simplified model is an
increase of the apex angle at the Br position of the molecule. Between ambient and
12 kbar pressure this angle $\alpha$, defined in Figure \ref{fig:cluster}b,
increases from 42.5$^{\circ}$ to 42.75$^{\circ}$. Since the Jahn-Teller axis of the
Mn$^{3+}$ coordination is close to the Mn$-$Br direction, pressure induces an inward
tilt of the three Mn$^{3+}$ anisotropy axes, thus decreasing the cluster anisotropy.
The value of the axial anisotropy parameter of the cluster in the $S = \frac{9}{2}$
ground state can be expressed as \cite{Gatteschi_Mn4}:

\begin{equation}\label{eq:D}
D_{\rm cluster}=\frac{105}{484}D_{\rm
Mn^{3+}}(3\cos^{2}\alpha-1)+\frac{35}{121}D_{33}-\frac{7}{44}D_{34}
\end{equation}

where $D_{\rm Mn^{3+}}$ is the single ion \textit{D} parameter of the Mn$^{3+}$,
$D_{33}$ and $D_{34}$ are magnetic dipole-dipole interaction terms between $\rm
Mn^{3+}-Mn^{3+}$ and $\rm Mn^{3+}-Mn^{4+}$, respectively. These latter two terms in
Eq.\ \ref{eq:D} can be calculated \cite{Gatteschi_Mn4}, they are typically an order
of magnitude smaller than the experimental \textit{D}, and their sum is practically
pressure independent, see Table \ref{table:calc}. We can thus definitely rule out
that the observed reduction of $|D_{\rm cluster}|$ with pressure is due to a change
in the dipole-dipole interaction. On the other hand, the $(3\cos^{2}\alpha-1)$
factor in the first term of Eq.\ \ref{eq:D} has a significant effect. With the
estimated increase of $\alpha$ by 0.25$^{\circ}$ at 12 kbar and taking $D_{\rm
Mn^{3+}}$ as pressure independent we calculate a 2.1\% reduction of the $|D_{\rm
cluster}|$ value at 12 kbar. This is to be compared with the reduction of 3.8\%
derived experimentally. We note that for $\alpha$ = 42.5$^{\circ}$ in Mn$_{4}$ the
function $(3\cos^{2}\alpha -1)$ is highly susceptible to minute changes of $\alpha$.
Since our compressibility model is rather crude, the estimated pressure dependence
of $\alpha$ has a relatively large uncertainty, which is amplified for the factor
$(3\cos^{2}\alpha -1)$. We therefore feel confident that we have identified the
principal structural element in Mn$_{4}$, which leads to a decrease of the axial
anisotropy under hydrostatic pressure. The compression of the apex with the
resulting increase of $\alpha$ is schematically represented in Figure
\ref{fig:cluster}b by the grey Br position.

\begin{table*}
\begin{tabular}{cccccc}
\hline \vspace{-8pt}\\
& $D_{\rm cluster}$ (meV) & $D_{\rm dd}$ (meV) & $D_{\rm Mn^{3+}}$ (meV) & $\alpha$ ($^{\circ}$) &$(3\cos^{2}\alpha -1)$ \\
  X =  & observed & calculated$^{b)}$ & calculated & & \vspace{2pt}\\
\hline \vspace{-8pt}\\
Br, ambient & -0.0627(1) & 0.0082 & -0.52 & 42.5(2)$^{c)}$ & 0.63 \\
Br, 12 kbar & -0.0603(3) & 0.0083 & -0.52$^{d)}$ & 42.75$^{e)}$ & 0.62 \\
Cl$^{a)}$ & -0.0656 & 0.0085 & -0.68 & 45.1(1)$^{c)}$ \vspace{2pt} & 0.50\\
\hline \vspace{-8pt}\\
\multicolumn{6}{l}{{\footnotesize$^{a)}$from Ref.\ \cite{Hanspeter}}}\\
\multicolumn{6}{l}{{\footnotesize$^{b)}$calculated using Ref.\ \cite{Gatteschi_Mn4}}}\\
\multicolumn{6}{l}{{\footnotesize$^{c)}$from Ref.\ \cite{Structure}}}\\
\multicolumn{6}{l}{{\footnotesize$^{d)}$assumed to be pressure independent}}\\
\multicolumn{6}{l}{{\footnotesize$^{e)}$calculated as described in the text}}\\
\end{tabular}
\caption{\label{table:calc} $D_{\rm cluster}$ values of
$\rm[Mn_{4}O_{3}X(OAc)_{3}(dbm)_{3}]$ (X $=$ Cl, Br) determined by INS. $D_{\rm dd}$
is the calculated sum of the two dipole-dipole contributions $D_{33}$ and $D_{34}$
in Eq.\ \ref{eq:D}. $D_{\rm Mn^{3+}}$ is the single-ion \textit{D} parameter
calculated from $D_{\rm cluster}$ using Eq.\ \ref{eq:D}. $\alpha$ is the angle at
the apex of the cluster defined in Figure \ref{fig:cluster}b, and $(3\cos^{2}\alpha
-1)$ is the factor in the first term of Eq.\ \ref{eq:D}.}
\end{table*}

Chemical variation is another way of tuning the cluster anisotropy. In Table
\ref{table:calc} we compare the effect of hydrostatic pressure on the Mn$_{4}$Br
compound with a chemical substitution of Br by Cl at ambient pressure. While
physical pressure mainly affects the $(3\cos^{2}\alpha-1)$ factor in the first term
of Eq.\ \ref{eq:D}, substitution of Br by Cl strongly increases the value of the
negative single-ion anisotropy parameter $D_{\rm Mn^{3+}}$ and thus more than
compensates for the decrease of the $(3\cos^{2}\alpha-1)$ factor. The result of the
chemical substitution from Br to Cl is thus an increase of $|D_{\rm cluster}|$,
while hydrostatic pressure of 12 kbar reduces it by about the same amount.

In conclusion, we have presented the first direct determination of the pressure
dependence of the axial anisotropy splitting in a SMM. Hydrostatic pressure of 12
kbar reduces the energy barrier between plus and minus spins by 3.8\%. The reduction
mainly results from a tilting of the single ion anisotropy axes of Mn$^{3+}$ under
pressure. Very recent INS experiments on Mn$_{12}$-acetate show a very small
increase of $|D|$ with increasing hydrostatic pressure. This different behavior
confirms our conclusion that the pressure dependence is a property determined by the
specific structure of a SMM molecule.

\begin{acknowledgments}
We thank Oliver Waldmann and Graham Carver for fruitful discussions. This work was
financially supported by the Swiss National Science Foundation (NFP 47) and the
European Union (TMR Molnanomag, No: HPRN-CT-1999-00012).
\end{acknowledgments}

\end{document}